\documentclass[aps,pra,preprint,groupedaddress]{revtex4-2}

\usepackage{amsmath}
\usepackage{amssymb}
\usepackage{graphicx} 
\usepackage{bm} 
\usepackage{hyperref} 

\usepackage{xcolor} 

\begin{document}

\title{
Information-Geometric Quantum Process Tomography in Unital Open Single-Qubit Dynamics}

\author{T. Koide}
\affiliation{Instituto de F\'{\i}sica, Universidade Federal do Rio de Janeiro, 
21941-972, Rio de Janeiro, RJ, Brazil}
\author{A. van de Venn}
\affiliation{Institute of Physics, University of Tartu, W. Ostwaldi 1, 50411 Tartu, Estonia}


\begin{abstract}
We derive an exact information-geometric inequality valid for both Markovian and non-Markovian mixed-state dynamics. This inequality saturates into a strict equality for single qubits because they belong to the quantum exponential family. 
This identity enables a non-iterative linear regression approach to continuous-time quantum process tomography, 
which, provided the full-rank condition is satisfied, yields a unique global solution and avoids local minima traps typical of standard non-linear optimization.
Furthermore, as the formulation is inherently unconstrained, negative dissipation rates provide direct evidence of non-Markovianity. 
Numerical simulations of the unital diagonal Bloch generator derived from the Gorini-Kossakowski-Sudarshan-Lindblad (GKSL) master equation demonstrate the validity of this geometric estimator and highlight the necessity of error mitigation near the pure-state boundary where the inverse metric becomes singular.
\end{abstract}

\maketitle


\section{Introduction}

Information-theoretic approaches are fundamental in characterizing correlations, entanglement, and the complexity of quantum systems \cite{nielsenbook,hayashibook,wildbook,watrousbook}. 
Among these, information geometry offers a powerful perspective by employing geometric concepts such as metrics, curvature, and geodesics to analyze statistical models \cite{Amari1985, AmariNagaoka2000, Ay2017}. 
The utility of information geometry is most profound when the underlying probability distributions (or density matrix operators) belong to an exponential family, where geometric structures such as metrics and affine connections are well-established at least in the classical case.

In general many-body systems, such as those described within the framework of Mori's projection operator method \cite{Mori1965}, 
observables are typically limited to macroscopic conserved densities (often called gross variables), which makes it difficult or impossible to uniquely determine the full microscopic state from observations.
This situation is different from few-body systems.
A typical example is a single-spin system, where the density matrix operators inherently form a quantum exponential family whose sufficient statistics are given by the Pauli matrices.
Crucially, this geometric structure of the quantum exponential family holds regardless of the specific details of the underlying dynamics, whether Markovian or non-Markovian, provided the time evolution is smooth and strictly confined to the interior of the state space (i.e., mixed states). 
We derive an information-geometric identity that is strictly satisfied by any single-qubit system. 
We establish this identity based on the general principles of information geometry, specifically by projecting the time-evolution operator onto the subspace spanned by the fluctuations of relevant observables. 
Theoretically, this identity can be viewed as a geometric derivation of Thermodynamic Speed Limits (TSLs), which is a topic actively debated in recent quantum and stochastic thermodynamics \cite{ItoDechant2020,Ito2018,Shiraishi2018,Nicholson2020,NishiyamaHasegawa2025,Funo2019,Hasegawa2020,Pires2016,Vu2023}. 
While TSLs are typically formulated as inequalities bounding the rate of state evolution, our geometric projection allows us to refine these bounds into a strict equality for qubits.

Beyond its theoretical significance, this geometric identity provides a practical advantage for parameter estimation in open quantum systems, specifically in continuous-time quantum process tomography.
Standard quantum process tomography typically relies on maximum likelihood estimation (MLE) \cite{fiurasek2001,sacchi2001,Childs2001,Jezek2003,Altepeter2003,Obrien2004,Ziman2005,Mohseni2008,Anis2012,Merkel2013,Samach2022}. 
This conventional approach requires solving complex non-linear optimization problems, which frequently lead to difficulties associated with local minima.
Alternatively, linear regression offers a highly tractable and computationally efficient approach. 
This method is utilized in quantum state tomography \cite{qi2013,hou2016,qi2017}, but its application to process tomography remains quite limited. 
For example, Ref.\ \cite{Yu2018} successfully applies linear regression estimation, but this approach restricts its target to closed quantum systems and requires subsequent matrix decompositions to identify the Hamiltonian. 
We also note that Ref.\ \cite{Dobrynin2025} employs MLE and considers a short-time linearization of the likelihood. 
This approach can be regarded as a form of linear regression, but it requires constrained optimization, such as projected gradient descent, to enforce the physical positivity of the parameters.

In this paper, we resolve these issues by utilizing our exact geometric identity to construct a natural loss function for the linear regression approach. 
Unlike classical probability distributions, where the Fisher information metric is typically regarded as unique, the quantum state space admits a family of monotone metrics \cite{petz1996}. 
While the symmetric logarithmic derivative (SLD) metric is standard in quantum estimation and defines the precision limits for parameter identification \cite{helstrom1976, Paris2009}, our geometric framework specifically exploits the Bogoliubov-Kubo-Mori (BKM) metric \cite{AmariNagaoka2000}.
Because this metric uniquely corresponds to the Hessian of the potential in the exponential family, it perfectly and exactly linearizes the inverse problem. 
The application of our abstract geometric identity allows us to develop a non-iterative method to simultaneously extract the Hamiltonian and dissipation parameters of the unital Gorini-Kossakowski-Sudarshan-Lindblad (GKSL) master equation directly from time-series data.

Furthermore, our formulation is inherently unconstrained and imposes no artificial boundary constraints on the physical parameter space, avoiding the need for algorithms like projected gradient descent.
If the estimation yields negative dissipation rates from observed time-series data, this result provides direct, regression-based evidence of non-Markovian dynamics. 
Thus, our approach serves as a robust and natural test to evaluate the validity of the Markovian assumption in the GKSL framework.

This paper is organized as follows. 
In Sec.\ \ref{sec:inequalities}, we establish the theoretical foundation by deriving universal information-geometric inequalities for general quantum and classical systems, and discuss their physical interpretations. 
In Sec.\ \ref{sec:identity}, we demonstrate that this inequality reduces to an exact geometric identity for single-qubit systems and apply it to the unital GKSL equation. 
Building on this strict identity, Sec.\ \ref{sec:parameter} introduces a practical parameter estimation method for the unital GKSL equation based on the linear regression approach. 
In Sec.\ \ref{sec:numerical_simulation}, we demonstrate the validity and noise-robustness of our method through numerical simulations. 
Finally, Sec.\ \ref{sec:summary} is devoted to discussions and concluding remarks.

\section{Universal Information-Geometric Inequalities} \label{sec:inequalities}

In this section, we present the theoretical foundation of our approach. 
While the fundamental mathematical inequalities discussed here are connected to existing concepts such as the Cram\'er-Rao bound and TSLs, our primary objective is to formulate them through a multidimensional and unified information-geometric projection.
The central result for our subsequent analysis of qubit dynamics is the quantum information-geometric inequality derived in Sec.~\ref{sec:quantum_bounds}, which explicitly isolates the information loss as a residual score. 
To elucidate our framework in the context of the existing literature, we also discuss its classical analogue in Sec.~\ref{sec:classical_bounds} and provide physical interpretations in Sec.~\ref{sec:interpretations}.

\subsection{Quantum Systems}
\label{sec:quantum_bounds}

We first consider a quantum system described by a density matrix operator $\hat{\rho}(\bm{\theta}(t))$, which depends on a set of $N$ time-dependent parameters $\bm{\theta}(t) = ( \theta^1 (t), \theta^2 (t), \cdots, \theta^N (t))$. 
The expectation value of an arbitrary time-independent operator $\hat{A}$ is given by:
\begin{equation}
\langle \hat{A} \rangle = \mathrm{Tr}[\hat{\rho}(\bm{\theta}(t))\hat{A}]\, .
\end{equation}
We assume that the density matrix operator is strictly positive (i.e., full-rank, having no zero eigenvalues). This assumption is necessary because, for pure states, the logarithm $\ln \hat{\rho}$ diverges, making the score operator ill-defined. Within this strictly positive domain, the density matrix operator can be written as $\hat{\rho}(\bm{\theta}(t)) = e^{\hat{H}(\bm{\theta}(t))}$ with a self-adjoint operator $\hat{H}(\bm{\theta}(t))$. The time derivative of the density matrix operator is expressed using the total score operator $\hat{\mathbb{L}}$:
\begin{equation}
\frac{d\hat{\rho}}{dt} = \int_{0}^{1} ds \, \hat{\rho}^{s} \hat{\mathbb{L}} \hat{\rho}^{1-s} \, ,
\end{equation}
where the total score operator is defined as the time derivative of the logarithm of the density matrix operator:
\begin{equation}
\hat{\mathbb{L}} :=  \frac{d}{dt} \ln \hat{\rho}(\bm{\theta}(t))  = \frac{d}{dt}\hat{H}(\bm{\theta}(t)) = \dot{\theta}^{\mu}\partial_{\mu}\hat{H}(\bm{\theta}(t)) \, ,
\end{equation}
with $\partial_\mu = \partial/\partial \theta^{\mu}$, and Greek indices running from $1$ to $N$.

Substituting this into the time derivative of the expectation value yields:
\begin{equation}
\frac{d}{dt}\langle \hat{A} \rangle 
= \mathrm{Tr}\left[\frac{d\hat{\rho}}{dt}\hat{A}\right] 
= \langle \hat{A}, \hat{\mathbb{L}} \rangle_{cc} 
= \langle \delta\hat{A}, \hat{\mathbb{L}} \rangle_{cc} \, .
\end{equation}
Here, we introduce the Kubo-Mori canonical correlation for two operators $\hat{X}$ and $\hat{Y}$ \cite{Mori1965}:
\begin{equation}
\langle \hat{X}, \hat{Y} \rangle_{cc} 
:= \int_{0}^{1} ds \, \mathrm{Tr}[\hat{\rho}^{s}\hat{X}\hat{\rho}^{1-s}\hat{Y}] \, .
\end{equation}
This canonical correlation satisfies the axioms of an inner product, and thus the space of operators equipped with $\langle \cdot, \cdot \rangle_{cc}$ forms a Hilbert space.
In the last equality, $\delta\hat{A} := \hat{A} - \langle \hat{A} \rangle$ represents the fluctuation of $\hat{A}$, and we utilized the property $\langle \hat{I}, \hat{\mathbb{L}} \rangle_{cc} = \mathrm{Tr}[d\hat{\rho}/dt] = 0$.

The geometry of the parameter space is characterized by the BKM metric $g_{\mu \nu}(\bm{\theta})$, defined as \cite{AmariNagaoka2000}:
\begin{equation}
\label{eq:general_metric_def}
g_{\mu \nu}(\bm{\theta}) := \langle \partial_{\mu}\hat{H}, \partial_{\nu}\hat{H} \rangle_{cc} \, .
\end{equation}
The interpretation of $g_{\mu\nu}(\bm{\theta})$ as a metric measuring the distinguishability of quantum states will be detailed later. 
The squared norm of the total score operator thus corresponds to
\begin{equation}
\label{eqn:LL-g}
\langle \hat{\mathbb{L}}, \hat{\mathbb{L}} \rangle_{cc} = g_{\mu \nu}(\bm{\theta})\dot{\theta}^{\mu}\dot{\theta}^{\nu} \, .
\end{equation}

Let us now introduce a set of $n\, (\le N)$ relevant observables $\hat{F} = (\hat{F}_{1}, \dots, \hat{F}_{n})^{T}$, defining the covariance matrix $\Xi (\bm{\theta})$ for their fluctuations:
\begin{equation}
\Xi_{ij} (\bm{\theta}) := \langle \delta\hat{F}_{i}, \delta\hat{F}_{j} \rangle_{cc} \, .
\end{equation}
We approximately express the total score operator $\hat{\mathbb{L}}$ by projecting it onto the subspace spanned by these fluctuations $\{\delta\hat{F}_{k}\}$:
\begin{equation}
    \tilde{\mathbb{L}} = C^{i}\delta\hat{F}_{i} \, ,
\end{equation}
where the Latin index runs from $1$ to $n (\le N)$.
The coefficients $C^{i}$ are determined by the orthogonality condition, 
\begin{equation}
 \langle \hat{e}, \delta\hat{F}_{i} \rangle_{cc} = 0 \, ,  
\end{equation}
where the residual score is defined by $\hat{e} :=\hat{\mathbb{L}} - \tilde{\mathbb{L}}$. This yields the following relation:
\begin{equation}
C = \Xi^{-1} (\bm{\theta})\frac{d}{dt}\langle \hat{F} \rangle \, ,
\end{equation}
where $C := (C^1,\cdots, C^n)^T$, $[\Xi]_{ij} := \Xi_{ij}$, and $\langle \hat{F} \rangle := (\langle \hat{F}^1 \rangle,\cdots, \langle \hat{F}^n \rangle)^T$.

Based on the Pythagorean theorem in the operator space defined by the canonical correlation, we have the orthogonal decomposition:
\begin{equation}
\langle \hat{\mathbb{L}}, \hat{\mathbb{L}} \rangle_{cc} = \langle \tilde{\mathbb{L}}, \tilde{\mathbb{L}} \rangle_{cc} + \langle \hat{e}, \hat{e} \rangle_{cc} \ge \langle \tilde{\mathbb{L}}, \tilde{\mathbb{L}} \rangle_{cc} \, ,
\end{equation}
owing to the non-negativity of the norm $\langle \hat{e}, \hat{e} \rangle_{cc} \ge 0$. 
Substituting the explicit forms into this relation, we arrive at the Quantum Information-Geometric Inequality:
\begin{equation}
\label{eq:general_inequality}
\left(\frac{d}{dt}\langle \hat{F} \rangle\right)^{T} \Xi^{-1} (\bm{\theta}) \frac{d}{dt}\langle \hat{F} \rangle \le g_{\mu \nu}(\bm{\theta})\dot{\theta}^{\mu}\dot{\theta}^{\nu} \, .
\end{equation}
The equality holds if and only if the relevant observables $\hat{F}$ completely capture the time evolution of the system (i.e., the residual $\hat{e}$ vanishes, meaning $\hat{F}$ constitutes sufficient statistics).
Crucially, this inequality is derived solely from the geometric projection in the operator space and does not depend on the specific details of the system's dynamics. It is universally satisfied for any Markovian or non-Markovian time evolution.

Finally, it is worth noting that approximately replacing the time evolution using a limited set of observables aligns with the philosophy of Mori's projection operator formalism \cite{Mori1965}. 
In that framework, the projection captures the coarse-grained dynamics of relevant variables, while the orthogonal component which corresponds to our residual score $\hat{e}$ is interpreted as the fluctuating (random) force. Dissipation is then expressed as the correlation of these fluctuating forces through the fluctuation-dissipation theorem of the second kind. 
Physically, the left-hand side of Eq.~(\ref{eq:general_inequality}) represents the ``coarse-grained" speed of the system observed through $\hat{F}$, while the right-hand side represents the ``microscopic (exact)" information-geometric speed. 
The inequality implies that coarse-graining always underestimates the intrinsic speed of state evolution.

We introduce the relative entropy as a measure of the distance between two states $\hat{\rho}(\bm{\theta}(t))$ and $\hat{\rho}(\bm{\theta}(t+dt))$:
\begin{equation}
\label{eq:rel_entropy_def}
D_{RE}[\hat{\rho}(\bm{\theta}(t)) || \hat{\rho}(\bm{\theta}(t+dt))] 
:= \mathrm{Tr}\left[ \hat{\rho}(\bm{\theta}(t)) (\ln \hat{\rho}(\bm{\theta}(t)) - \ln \hat{\rho}(\bm{\theta}(t+dt))) \right] \, .
\end{equation}
Expanding the relative entropy up to the second order in $dt$, and using the geometric constraints $\langle \hat{\mathbb{L}} \rangle = 0$ and $\langle \dot{\hat{\mathbb{L}}} \rangle = - \langle \hat{\mathbb{L}}, \hat{\mathbb{L}} \rangle_{cc}$, we obtain:
\begin{equation}
\label{eq:speed_rel_entropy}
g_{\mu \nu}(\bm{\theta})\dot{\theta}^{\mu}\dot{\theta}^{\nu} = \lim_{dt\to 0} \frac{2 D_{RE}[\hat{\rho}(\bm{\theta}(t)) || \hat{\rho}(\bm{\theta}(t+dt))]}{(dt)^{2}} \, .
\end{equation}
Consequently, the general inequality (\ref{eq:general_inequality}) can be rewritten in terms of the relative entropy as:
\begin{equation}
\label{eqn:UQIGI-2}
\left(\frac{d}{dt}\langle \hat{F} \rangle\right)^{T} \Xi^{-1} (\bm{\theta})\frac{d}{dt}\langle \hat{F} \rangle \le \lim_{dt\to 0} \frac{2 D_{RE}[\hat{\rho}(\bm{\theta}(t)) || \hat{\rho}(\bm{\theta}(t+dt))]}{(dt)^{2}} \, .
\end{equation}
This relation establishes that the coarse-grained evolution rate is fundamentally bounded by the intrinsic information-geometric rate.

\subsection{Classical Systems}
\label{sec:classical_bounds}

The universal inequality (\ref{eqn:UQIGI-2}) derived for quantum systems has a direct counterpart in classical stochastic systems.
Let us consider a probability distribution $p(\bm{x}; \bm{\theta}(t))$ defined over a state $\bm{x}$, parametrized by $N$ time-dependent parameters $\bm{\theta}(t)=(\theta^{1}(t), \dots, \theta^N (t))$.

The geometry of the parameter space is governed by the classical Fisher information metric $g_{ij}(\bm{\theta})$, defined as:
\begin{equation}
g_{ij}(\bm{\theta}) = \int d\bm{x}\, p(\bm{x}; \bm{\theta}) \frac{\partial \ln p(\bm{x}; \bm{\theta})}{\partial \theta^i} \frac{\partial \ln p(\bm{x}; \bm{\theta})}{\partial \theta^j} 
:= \mathbb{E}_{\bm{\theta}}[(\partial_{i}\ln p)(\partial_{j}\ln p)]\, .
\end{equation}

We define the classical total score function $\mathbb{L}(\bm{x},t)$ as the logarithmic time derivative of the probability distribution:
\begin{equation}
\mathbb{L}(\bm{x},t) := \frac{d}{dt} \ln p(\bm{x}; \bm{\theta}(t)) = \dot{\theta}^i \partial_i \ln p(\bm{x}; \bm{\theta}(t)) \, .
\end{equation}
Note that $\mathbb{E}_{\bm{\theta}}[\mathbb{L}] = 0$ due to probability conservation.

Now, consider $n\, (\le N)$ relevant observables $\bm{f}=(f_{1}, \dots, f_{n})^{T}$. Let $\Xi(\bm{\theta})$ be the covariance matrix of their fluctuations $\delta f_{i} := f_{i} - \mathbb{E}_{\bm{\theta}}[f_{i}]$:
\begin{equation}
\Xi_{ij}(\bm{\theta}) := \mathbb{E}_{\bm{\theta}}[\delta f_i \delta f_j] \, .
\end{equation}

We project the total score function $\mathbb{L}$ onto the subspace spanned by these fluctuations:
\begin{equation}
\tilde{\mathbb{L}} = C^{i} \delta f_{i} \, .
\end{equation}
The coefficients $C^i$ are determined by the orthogonality condition $\mathbb{E}_{\bm{\theta}}[(\mathbb{L} - \tilde{\mathbb{L}})\delta f_j] = 0$, which yields
\begin{equation}
C = \Xi^{-1} (\bm{\theta})\frac{d}{dt}\mathbb{E}_{\bm{\theta}}[\bm{f}] \, .
\end{equation}

By applying the Pythagorean theorem $\mathbb{E}_{\bm{\theta}}[\mathbb{L}^2] = \mathbb{E}_{\bm{\theta}}[\tilde{\mathbb{L}}^2] + \mathbb{E}_{\bm{\theta}}[(\mathbb{L} - \tilde{\mathbb{L}})^2] \ge \mathbb{E}_{\bm{\theta}}[\tilde{\mathbb{L}}^2]$, we arrive at the classical inequality:
\begin{equation}
\left(\frac{d}{dt} \mathbb{E}_{\bm{\theta}}[ \bm{f} ]\right)^{T}\Xi^{-1}(\bm{\theta}) \frac{d}{dt}\mathbb{E}_{\bm{\theta}}[ \bm{f} ]
\le
g_{ij}(\bm{\theta})\dot{\theta}^{i}\dot{\theta}^{j}
\, .
\label{eq:classical_inequality}
\end{equation}
As is well-known in information geometry \cite{Amari1985, AmariNagaoka2000, Ay2017}, the right-hand side can be expressed in terms of the Kullback-Leibler (KL) divergence: 
\begin{equation}
  g_{ij}(\bm{\theta})\dot{\theta}^{i}\dot{\theta}^{j}
=
\lim_{dt\rightarrow0}\frac{2D_{KL}[p(\bm{x}; \bm{\theta}(t))||p(\bm{x}; \bm{\theta}(t+dt))]}{(dt)^{2}} \, ,  
\end{equation}
where
\begin{equation}
   D_{KL}[A(\bm{x})||B(\bm{x})]
   = 
   \int d \bm{x} \, A(\bm{x}) \ln \frac{A(\bm{x})}{B(\bm{x})} \, .
\end{equation}
This inequality formally states that the rate of change of macroscopic observables is bounded by the intrinsic information-geometric rate of the underlying probability distribution.

\subsection{Physical and Statistical Interpretations}
\label{sec:interpretations}

The inequalities derived above, Eqs.~(\ref{eqn:UQIGI-2}) and (\ref{eq:classical_inequality}), can be understood as the dynamical counterparts of the Cram\'er-Rao bound in estimation theory.
Let us consider estimating a single scalar parameter $\theta(t)$ using a single observable estimator $f(\bm{x})$ ($n=1$ in Eq.~(\ref{eq:classical_inequality})).
We assume that the estimator $f(\bm{x})$ is unbiased, which is mathematically defined by the condition $\mathbb{E}_{\theta}[f] = \theta$.
In this case, the time derivative of the expectation value is simply the speed of the parameter itself:
\begin{equation}
\frac{d}{dt}\mathbb{E}_{\theta}[f] = \dot{\theta}\, .
\end{equation}
The covariance matrix $\Xi$ then reduces to the variance of the estimator, $\Xi = \text{Var}(f)$, and the Fisher metric $g_{\mu\nu}\dot{\theta}^\mu\dot{\theta}^\nu$ simplifies to $I(\theta)\dot{\theta}^2$, where $I(\theta)$ is the Fisher information, $\mathbb{E}_{\theta}[(\partial_\theta \ln p(\theta))^2 ]$.
Substituting these into the inequality (\ref{eq:classical_inequality}), we obtain:
\begin{equation}
\frac{(\dot{\theta})^2}{\text{Var}(f)} \le I(\theta) (\dot{\theta})^2 \, .
\end{equation}
Assuming $\dot{\theta} \neq 0$, this immediately yields the standard Cram\'er-Rao bound:
\begin{equation}
\text{Var}(f) \ge \frac{1}{I(\theta)} \, .
\end{equation}
This result confirms that our information-geometric inequality is linked to the precision of parameter estimation. 
The ``squared norm" on the right-hand side of our universal inequalities sets the fundamental bound on the rate at which any physical observable can evolve.
While the standard Cram\'er-Rao bound is typically considered in a static context, it reflects the intrinsic geometry of the statistical manifold within the framework of information geometry. 
Even when a state changes over time, the dynamics are fundamentally constrained by this underlying geometry. 
In this sense, the reproduction of the Cram\'{e}r-Rao bound in the present dynamical situation stands to reason.

Furthermore, these inequalities are connected to the concept of Thermodynamic Speed Limits (TSLs), a subject of extensive recent study \cite{ItoDechant2020,Ito2018,Shiraishi2018,Nicholson2020,NishiyamaHasegawa2025,Funo2019,Hasegawa2020,Pires2016,Vu2023}. 
In the classical regime, our derived bound, Eq.~(\ref{eq:classical_inequality}), is formally equivalent to the speed limits based on Fisher information discussed by Ito and Dechant \cite{Ito2018, ItoDechant2020}, and is closely related to the limits imposed by entropy production and dynamical activity derived by Shiraishi \textit{et al.} \cite{Shiraishi2018} and Nicholson \textit{et al.} \cite{Nicholson2020}. 
It is worth emphasizing that while TSLs are often discussed by taking the Cram\'er-Rao inequality as a phenomenological starting point, our approach derives them explicitly from a geometric identity based on mathematical projection.
This suggests that these dynamic bounds are intrinsic geometric features of state evolution, existing independently of any specific estimation context.

While the geometric interpretation of information is uniquely established for classical systems, the quantum case admits a family of metrics, as mentioned in the Introduction. 
Standard quantum estimation theory typically employs the SLD metric to define the theoretical bound on estimation precision \cite{AmariNagaoka2000,petz1996,helstrom1976}. 
In contrast, our geometric projection method naturally yields a Cram\'er-Rao-type inequality characterized by the BKM metric.
We adopt this BKM formulation for its unique structural properties rather than for achieving the tightest possible estimation bound.
As shown by Scandi and Perarnau-Llobet \cite{Scandi2019} (see also Ref.~\cite{Dechant2022}), the BKM metric is the natural geometric measure for describing thermodynamic length and dissipation in open quantum systems within the linear response regime. 
Crucially for our practical objective, the BKM metric corresponds to the Hessian of the potential in the exponential family. This mathematical property uniquely simplifies the complex inverse problem of parameter estimation into a highly tractable linear regression, as will be demonstrated in Sec.~\ref{sec:LRE}. 
Unlike the BKM metric, the SLD metric lacks the algebraic structure required to linearize the inverse problem. 
Therefore, any attempt to use the SLD metric would entangle the equations in a highly non-linear optimization problem, making the analytical and numerical extraction of parameters vastly more difficult.

We confirmed that our quantum inequality \eqref{eq:general_inequality} generalizes the result by Lacerda et al. \cite{Lacerda}, who extended the speed limits on observables by Garc\'ia-Pintos et al. \cite{Pintos} to general monotone metrics.
In the single-observable limit, their inequality (28) mathematically coincides with our result when the BKM metric is chosen as the monotone metric.
However, while their formulation is restricted to a single scalar observable, our multidimensional approach accommodates multiple variables simultaneously and correctly incorporates their cross-correlations through the covariance matrix $\Xi^{-1}$.
Furthermore, our projection-based derivation explicitly clarifies the algebraic condition under which the inequality saturates into a strict identity, which occurs when the projection residual $\hat{e}$ is zero.
In contrast, Garc\'ia-Pintos et al. and Lacerda et al. bypass the explicit algebraic conditions for a zero residual, and instead relate the saturation of the dynamic bound to the path action along a geodesic to minimize thermodynamic dissipation.

If a physical system inherently satisfies the equality condition (i.e., the residual vanishes unconditionally, $\hat{e}=0$), this identity enables practical parameter estimation. In the subsequent sections, we demonstrate this for an open single-qubit system and develop a robust estimation method.

\section{Exact Geometric Identity for Single-Qubit Systems} \label{sec:identity}

In this section, we apply our general framework to a single-qubit system. 
While the relations derived in the previous section are inequalities due to the potential loss of information during the projection step, we show here that for a single qubit, this inequality becomes a strict identity. 
Because the Pauli matrices provide sufficient statistics for the quantum exponential family of a qubit, they capture all information about the state evolution without any loss.

\subsection{Derivation of the Universal Identity}

The state of a single qubit is conventionally described using the Bloch vector representation:
\begin{equation}
\label{eq:bloch_rep}
\hat{\rho} = \frac{1}{2}\left(\mathbb{I}_{2} + \sum_{\mu=1}^3 a_{\mu}\sigma_{\mu}\right) \, ,
\end{equation}
where $a_\mu = \mathrm{Tr}[\hat{\rho}\sigma_\mu]$ corresponds to the expectation values of the Pauli matrices. The vector $\bm{a} = (a_1, a_2, a_3)^T$ resides within the unit sphere in $\mathbb{R}^3$, which is defined by the standard Euclidean norm $|\bm{a}|^2 := \sum_{\mu=1}^3 (a_\mu)^2 \le 1$.
This space is naturally equipped with the standard Euclidean metric $\delta_{\mu\nu}$. 
In the following discussion, whenever we use vector notation such as $|\bm{a}|$ or $\bm{a} \times \bm{b}$, it implies operations with respect to this Euclidean geometry.

From the perspective of information geometry, however, the set of mixed states of a single qubit forms a smooth statistical manifold equipped with the BKM metric $g_{\mu\nu}$. This metric quantifies state distinguishability and governs the information-geometric cost of transitions. 
The qubit state belongs to a quantum exponential family where the density matrix operator in Eq.~(\ref{eq:bloch_rep}) can be reexpressed in the following canonical exponential form:
\begin{equation}
\label{eq:exp_family_rep}
\hat{\rho}(\bm{\theta}) = \exp\left(\theta^{\mu}\sigma_{\mu} - \psi(\bm{\theta}) \right) \, ,
\end{equation}
where $\theta^\mu$ are the natural parameters given by
\begin{equation}
\theta^\mu = \frac{\text{arctanh}(|\bm{a}|)}{|\bm{a}|} a_\mu \, ,
\end{equation}
and the potential $\psi(\bm{\theta})$ is determined from the normalization condition of the density matrix operator, $\mathrm{Tr}[\hat{\rho}]=1$, yielding
\begin{equation}
\psi(\bm{\theta}) = \ln \mathrm{Tr} \left[ \exp \left( \theta^{\mu}\sigma_{\mu} \right) \right] \, . 
\end{equation}

Within the framework of the quantum exponential family, the Pauli matrices $\boldsymbol{\sigma}$ constitute the sufficient statistics of the distribution, where $N = n = 3$. 
Consequently, the relevant observables $\hat{F}=\boldsymbol{\sigma}$ capture all information about the state evolution. The residual score $\hat{e}$ in the derivation of the general inequality (\ref{eq:general_inequality}) vanishes identically because the total score operator $\hat{\mathbb{L}}$ is completely contained in the space spanned by the fluctuation of the sufficient statistics defined by $\sigma_\mu$. Thus, for any single-qubit system, whether the dynamics are Markovian or non-Markovian, the inequality is replaced by a strict equality:
\begin{equation}
\label{eq:strict_equality}
\left(\frac{d}{dt}\langle \boldsymbol{\sigma} \rangle\right)^{T} \Xi^{-1} (\bm{\theta})\frac{d}{dt}\langle \boldsymbol{\sigma} \rangle 
= 
 \lim_{dt\to 0} \frac{2 D_{RE}[\hat{\rho}(\bm{\theta}(t)) || \hat{\rho}(\bm{\theta}(t+dt))]}{(dt)^{2}}\, .
\end{equation}

In this case, the BKM metric $g_{\mu\nu}$, which coincides with the covariance matrix $\Xi_{\mu\nu}(\bm{\theta})$, can be calculated explicitly:
\begin{equation}
\label{eq:qubit_metric}
g_{\mu\nu}(\bm{\theta}) =\Xi_{\mu\nu} (\bm{\theta}) = \frac{|\bm{a}|}{\text{arctanh}(| \bm{a}|)} \left( \delta_{\mu\nu} - \frac{a_\mu a_\nu}{|\bm{a}|^2} \right) + (1-| \bm{a}|^2) \frac{a_\mu a_\nu}{|\bm{a}|^2} \, .
\end{equation}
This metric tensor fundamentally characterizes the geometry of the qubit state space. 
Using this explicit form, we arrive at the universal information-geometric equality for a single qubit:
\begin{equation}
\label{eq:qubit_identity_general1}
\frac{1}{1-|\bm{a}|^2}\left(\frac{d|\bm{a}|}{dt}\right)^2 + \frac{\text{arctanh}(| \bm{a}|)}{|\bm{a}|^3} |\bm{a} \times \dot{\bm{a}}|^2 = \lim_{dt\to 0} \frac{2 D_{RE}[\hat{\rho}(\bm{\theta}(t)) || \hat{\rho}(\bm{\theta}(t+dt))]}{(dt)^2} \, ,
\end{equation}
where we utilized the inverse matrix
\begin{equation}
g^{-1}_{\mu\nu}(\bm{\theta})
=
\Xi^{-1}_{\mu\nu} (\bm{\theta}) 
=  
\frac{\text{arctanh}(|\bm{a}|)}{|\bm{a}|} \left( \delta_{\mu\nu} - \frac{a_\mu a_\nu}{| \bm{a}|^2} \right) + \frac{1}{1-|\bm{a}|^2} \frac{a_\mu a_\nu}{|\bm{a}|^2} \, .
\end{equation}

The two terms on the left-hand side of Eq.~(\ref{eq:qubit_identity_general1}) have distinct physical meanings. The first term, proportional to $(d|\bm{a}|/dt)^2$, corresponds to the radial contribution associated with the change in purity. 
Conversely, the second term, proportional to $|\bm{a} \times \dot{\bm{a}}|^2$, represents the angular contribution associated with the rotation of the state.
Due to the divergent factor $1/(1-|\bm{a}|^2)$ near $|\bm{a}| \longrightarrow 1$, the first term indicates that nearly pure states are geometrically rigid and difficult to control precisely.

\subsection{Application to unital GKSL Dynamics} \label{sec:LRE}

Let us now apply this identity to a single-qubit system governed by the unital GKSL equation. We suppose that the system Hamiltonian is given by $\hat{H}=\bm{e}\cdot\bm{\sigma}$ with a constant vector $\bm{e}$. 
In general, the Bloch equation derived from the GKSL master equation can take a more general form that includes a non-diagonal dissipative matrix and a non-zero affine term 
(see Appendix \ref{app:affine_derivation}). 
In this paper, to simplify the numerical demonstration, we assume a simplified dynamics of the Bloch vector:
\begin{equation}
\label{eq:gksl_bloch}
\dot{\bm{a}} = 2\bm{e} \times \bm{a} - 2D_r \bm{a} \, ,
\end{equation}
where $D_r= \operatorname{diag}(d_1, d_2, d_3)$ is the dissipation constant matrix characterizing the interaction with the environment.
Here, the system simply relaxes toward the maximally mixed state, meaning $\bm{a} \to \bm{0}$.

Substituting this Euclidean equation of motion into the left-hand side of Eq.~(\ref{eq:qubit_identity_general1}), we obtain:
\begin{equation}
\label{eq:qubit_identity_gksl}
\frac{4(\bm{a}^T D_r \bm{a})^2}{|\bm{a}|^2(1-| \bm{a}|^2)} + \frac{4\text{arctanh}(| \bm{a}|)}{|\bm{a}|^3} \left| |\bm{a}|^2 \bm{e}_{\perp} - \bm{a}\times (D_r \bm{a}) \right|^2 = \lim_{dt\to 0} \frac{2 D_{RE}[\hat{\rho}(\bm{\theta}(t)) || \hat{\rho}(\bm{\theta}(t+dt))]}{(dt)^2},
\end{equation}
where we have defined the orthogonal component $\bm{e}_{\perp} = \bm{e} - \frac{(\bm{a}\cdot \bm{e})}{|\bm{a}|^2}\bm{a}$.

\section{Parameter Estimation Method for Unital GKSL Dynamics}
\label{sec:parameter}

In this section, we propose a method to estimate the Hamiltonian vector $\bm{e}$ and the dissipation matrix $D_r$ from the time-series data of the Bloch vector $\bm{a}(t)$, utilizing the information-geometric framework established above.


Let us consider an experimentally observed time series of the Bloch vector $\bm{a}(t)$. 
We assume that the system dynamics are governed by the unital GKSL equation, which is characterized by the Hamiltonian vector $\bm{e}$ and the dissipation matrix $D_r = \operatorname{diag}(d_1, d_2, d_3)$.
Ideally, this time evolution must strictly satisfy the information-geometric identities derived in the previous sections.
By equating the left-hand sides of Eqs.~(\ref{eq:qubit_identity_general1}) and (\ref{eq:qubit_identity_gksl}), we obtain the following relation linking the microscopic parameters to the observed kinematics:
\begin{align}
\lefteqn{\frac{4(\bm{a}^T D_r \bm{a})^2}{|\bm{a}|^2(1-|\bm{a}|^2)} + \frac{4 \operatorname{arctanh}(|\bm{a}|)}{|\bm{a}|^3} \left||\bm{a}|^2 \bm{e}_{\perp} - \bm{a} \times (D_r \bm{a})\right|^2} & \nonumber \\
&= \frac{1}{1-|\bm{a}|^2} \left(\frac{d|\bm{a}|}{dt}\right)^2 + \frac{\operatorname{arctanh}(|\bm{a}|)}{|\bm{a}|^3}|\bm{a} \times \dot{\bm{a}}_{\text{ob}}|^2 \, .
\label{eq:full_identity}
\end{align}
Here we denote an observed ``velocity" by $\dot{\bm{a}}_{\text{ob}}(t)$ for a given $\bm{a}(t)$.

We define the residual velocity vector $\Delta \bm{v}$ as the difference between the observed velocity $\dot{\bm{a}}_{\text{ob}}(t)$ and the theoretical velocity predicted by the GKSL model $\bm{v}_{\text{model}}(t)$:
\begin{equation}
    \Delta \bm{v}(t) := \dot{\bm{a}}_{\text{ob}}(t) - \bm{v}_{\text{model}}(t) \, ,
\end{equation}
where 
\begin{equation}
\bm{v}_{\text{model}} (t)= 2\bm{e} \times \bm{a} (t)- 2D_r \bm{a} (t) \, . 
\label{eqn:v_model}
\end{equation}

A crucial feature of the GKSL dynamics is that Eq.~(\ref{eqn:v_model}) is linear with respect to the unknown parameters. 
Let us define the parameter vector $\bm{p}$ containing the components of the Hamiltonian vector and the diagonal elements of the dissipation matrix:
\begin{equation}
    \bm{p} = (e_1, e_2, e_3, d_1, d_2, d_3)^T \, .
\end{equation}
The model velocity (\ref{eqn:v_model}) can then be expressed as a linear transformation of $\bm{p}$:
\begin{equation}
    \bm{v}_{\text{model}}(t) = H_t \bm{p} \, ,
\end{equation}
where the matrix $H_t \in \mathbb{R}^{3 \times 6}$ depends only on the instantaneous state $\bm{a}(t) = (a_1, a_2, a_3)^T$:
\begin{equation}
    H_t 
    = \begin{pmatrix}
        0 & 2a_3(t) & -2a_2(t) & -2a_1(t) & 0 & 0 \\
        -2a_3(t) & 0 & 2a_1(t) & 0 & -2a_2(t) & 0 \\
        2a_2(t) & -2a_1(t) & 0 & 0 & 0 & -2a_3(t)
    \end{pmatrix} \, .
\end{equation} 

In Eq.~(\ref{eq:full_identity}), only the right-hand side contains the velocity $\dot{\bm{a}}_{\text{ob}}(t)$. This velocity can be expressed as $\dot{\bm{a}}_{\text{ob}}(t) = \bm{v}_{\text{model}}(t) + \Delta \bm{v}(t)$, keeping in mind the relation $d|\bm{a}_{\text{ob}}|/dt = (\bm{a}_{\text{ob}} \cdot \dot{\bm{a}}_{\text{ob}}) / |\bm{a}_{\text{ob}}|$.
Equation~(\ref{eq:full_identity}) can thus be reexpressed as:
\begin{equation}
\frac{4(\bm{a}^T D_r \bm{a})^2}{|\bm{a}|^2(1-|\bm{a}|^2)} + \frac{4 \operatorname{arctanh}(|\bm{a}|)}{|\bm{a}|^3} \left||\bm{a}|^2 \bm{e}_{\perp} - \bm{a} \times (D_r \bm{a})\right|^2
=
(\bm{v}_{\text{model}} +  \Delta \bm{v})^T G^{-1}(\bm{a}) (\bm{v}_{\text{model}} +  \Delta \bm{v}) \, ,
\end{equation}
where $G^{-1}(\bm{a})$ is the matrix representation of the inverse BKM metric $g^{\mu\nu}(\bm{\theta})$.
In the ideal case without experimental noise ($\Delta \bm{v}(t)=0$), the left-hand side simply equals $\bm{v}_{\text{model}}^T G^{-1}(\bm{a}) \bm{v}_{\text{model}}$. Expanding the right-hand side then leads to the following relation:
\begin{align}
0 &= (\bm{v}_{\text{model}} + \Delta \bm{v})^T G^{-1}(\bm{a}) (\bm{v}_{\text{model}} + \Delta \bm{v}) - \bm{v}_{\text{model}}^T G^{-1}(\bm{a}) \bm{v}_{\text{model}} \nonumber \\
 &= 2 \bm{v}_{\text{model}}^T G^{-1}(\bm{a}) \Delta \bm{v} + \Delta \bm{v}^T G^{-1}(\bm{a}) \Delta \bm{v} \, .
    \label{eq:J_definition}
\end{align}
Therefore, the true parameters of the GKSL equation correspond to the minimum of a positive-definite quantity $\mathcal{J}_2$, which we choose as the loss function for our optimization:
\begin{equation}
    {\cal J}_2 :=\Delta \bm{v}^T G^{-1}(\bm{a}) \Delta \bm{v} \, .
\end{equation}
Because $G^{-1}(\bm{a})$ is positive-definite, finding parameters that yield $\mathcal{J}_2=0$ strictly guarantees $\Delta \bm{v} = \bm{0}$. Consequently, the first term of Eq.~(\ref{eq:J_definition}) also vanishes, and Eq.~(\ref{eq:full_identity}) is exactly satisfied. This yields the exact parameters of the GKSL equation that reproduce the ideal experimental data without any errors.

However, for a single time point $t$, the matrix $H_t$ is rank-deficient, meaning that the parameters cannot be uniquely determined from a single observation.
To uniquely identify all parameters, it is necessary to accumulate geometric information over the entire time evolution. 
Therefore, we define the total loss function $\mathcal{L}(\bm{p})$ by summing $\mathcal{J}_2(t)$ over all observed time points.
Substituting $\Delta \bm{v}(t) = \dot{\bm{a}}_{\text{ob}}(t) - H_t \bm{p}$ into the expression for $\mathcal{J}_2(t)$, the total loss function $\mathcal{L}(\bm{p})$ becomes:
\begin{equation}
    \mathcal{L}(\bm{p}) = \sum_{t} \mathcal{J}_2(t) = \sum_{t} (\dot{\bm{a}}_{\text{ob}}(t) - H_t \bm{p})^T G^{-1}(\bm{a}(t)) (\dot{\bm{a}}_{\text{ob}}(t) - H_t \bm{p}) \, .
\end{equation}
We then seek the parameters that satisfy the variational condition:
\begin{equation}
\left. \frac{\delta \mathcal{L} (\bm{p})}{\delta \bm{p}}\right|_{\bm{p}=\bm{p}^*} = 0 \, .
\end{equation}
This minimization of the loss function $\mathcal{L}$ with respect to $\bm{p}$ leads to the following system of linear equations:
\begin{equation}
    \sum_{t} H_t^T G^{-1}(\bm{a}(t)) H_t \bm{p}^* = \sum_{t} H_t^T G^{-1}(\bm{a}(t)) \dot{\bm{a}}_{\text{ob}}(t) \, .
\end{equation}
This can be written compactly as:
\begin{equation}
    \bm{p}^* = A^{-1}\bm{b} \, ,
\end{equation}
where the matrix $A$ and the vector $\bm{b}$ are obtained by accumulating data over the time series:
\begin{align}
    A &= \sum_{t} H_t^T G^{-1}(\bm{a}(t)) H_t \, , \\
    \bm{b} &= \sum_{t} H_t^T G^{-1}(\bm{a}(t)) \dot{\bm{a}}_{\text{ob}}(t) \, .
\end{align}

We emphasize that a single trajectory, or trajectories starting from certain symmetric initial states, may lack sufficient information to make the matrix $A$ full rank, leaving some parameters unidentifiable. 
To obtain a unique global solution and to enhance numerical stability, 
the use of time-series data from multiple trajectories starting from a set of linearly independent initial states is required.
Since the estimation problem is linear, these independent datasets can be simply aggregated into the summations defining $A$ and $\bm{b}$.
Consequently, the use of appropriate datasets is expected to guarantee the full-rank property of matrix $A$.

\section{Numerical Simulations} \label{sec:numerical_simulation}

\begin{figure}[h]
\begin{center}
\includegraphics[scale=0.3]{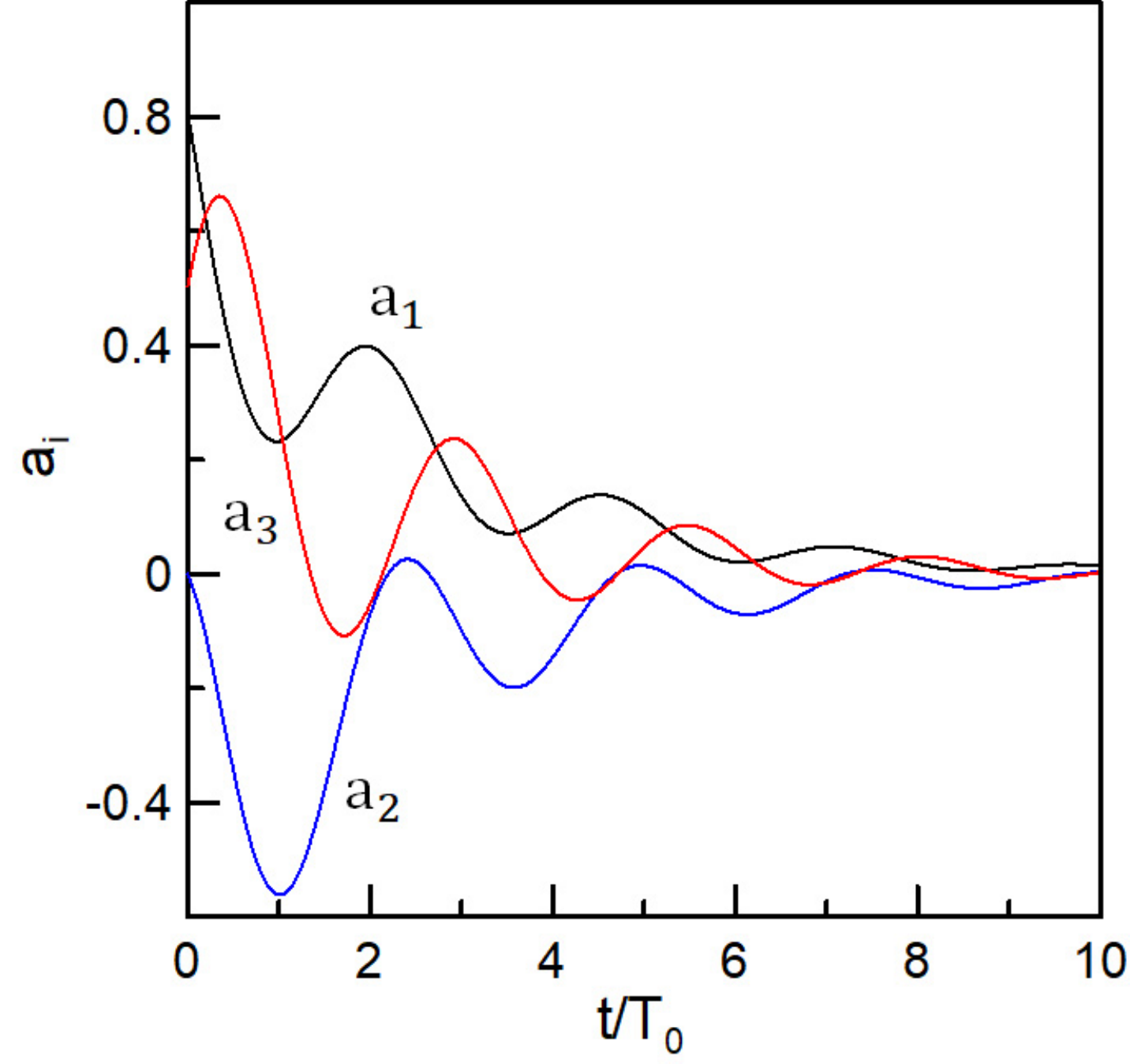}
\includegraphics[scale=0.3]{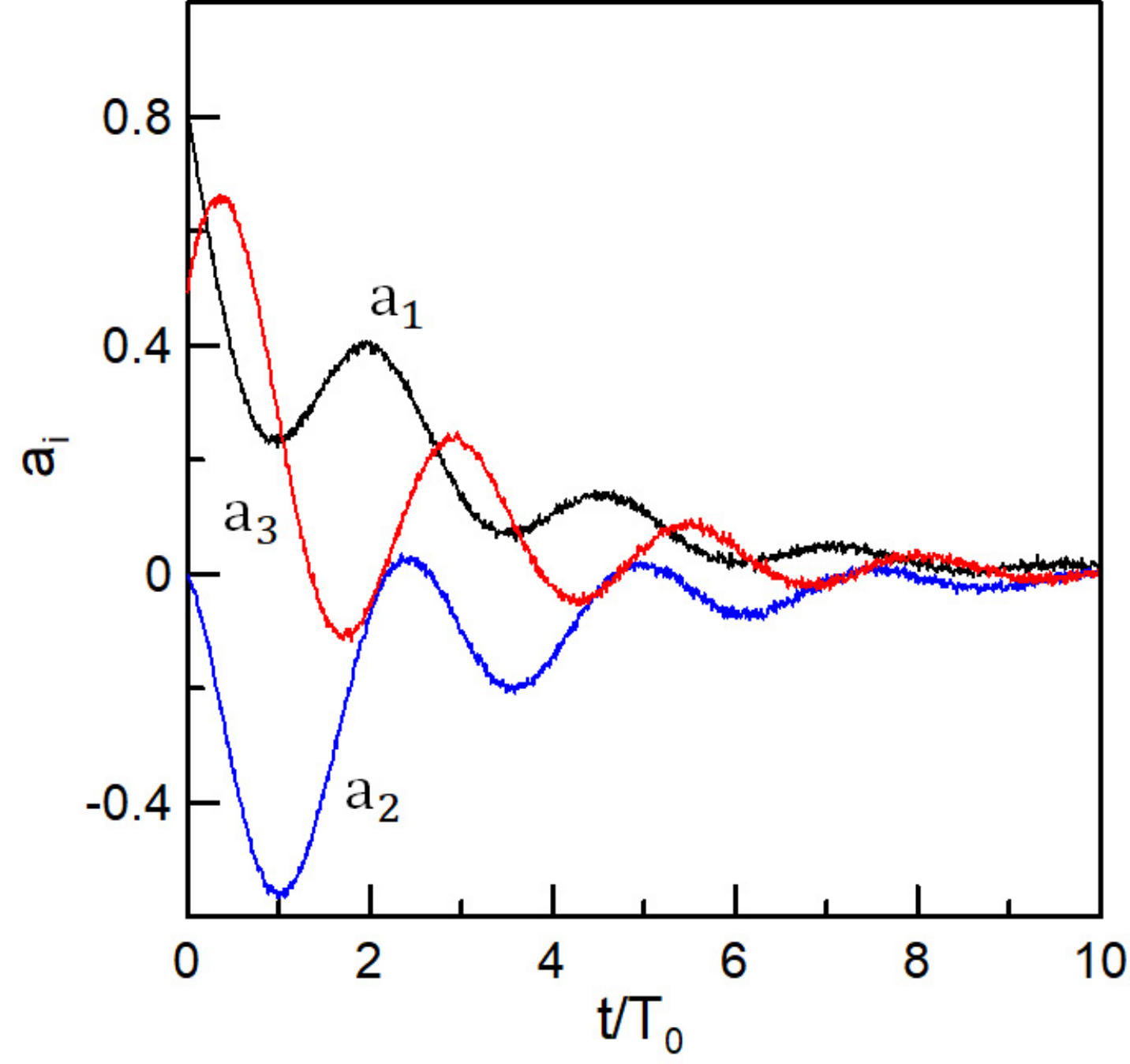}
\caption{
Time evolution of the Bloch vector components $a_i$. 
The horizontal axis represents the dimensionless time $t/T_0$, where $T_0$ is a constant with the dimension of time. 
The left panel shows ideal trajectories governed by the GKSL equation. 
The right panel shows trajectories perturbed by Gaussian white noise to mimic  experimental errors.
These results are obtained using the parameters $\bm{e} = (1.0, -0.6, 0.4)/T_0$ and $\bm{d} = (0.2, 0.3, 0.1)/T_0$, 
with the initial condition $\bm{a}(0) = (0.815, -0.007, 0.466)$.
}
\label{fig:bloch_vec}
\end{center}
\end{figure}

\begin{figure}[h]
\begin{center}
\includegraphics[scale=0.3]{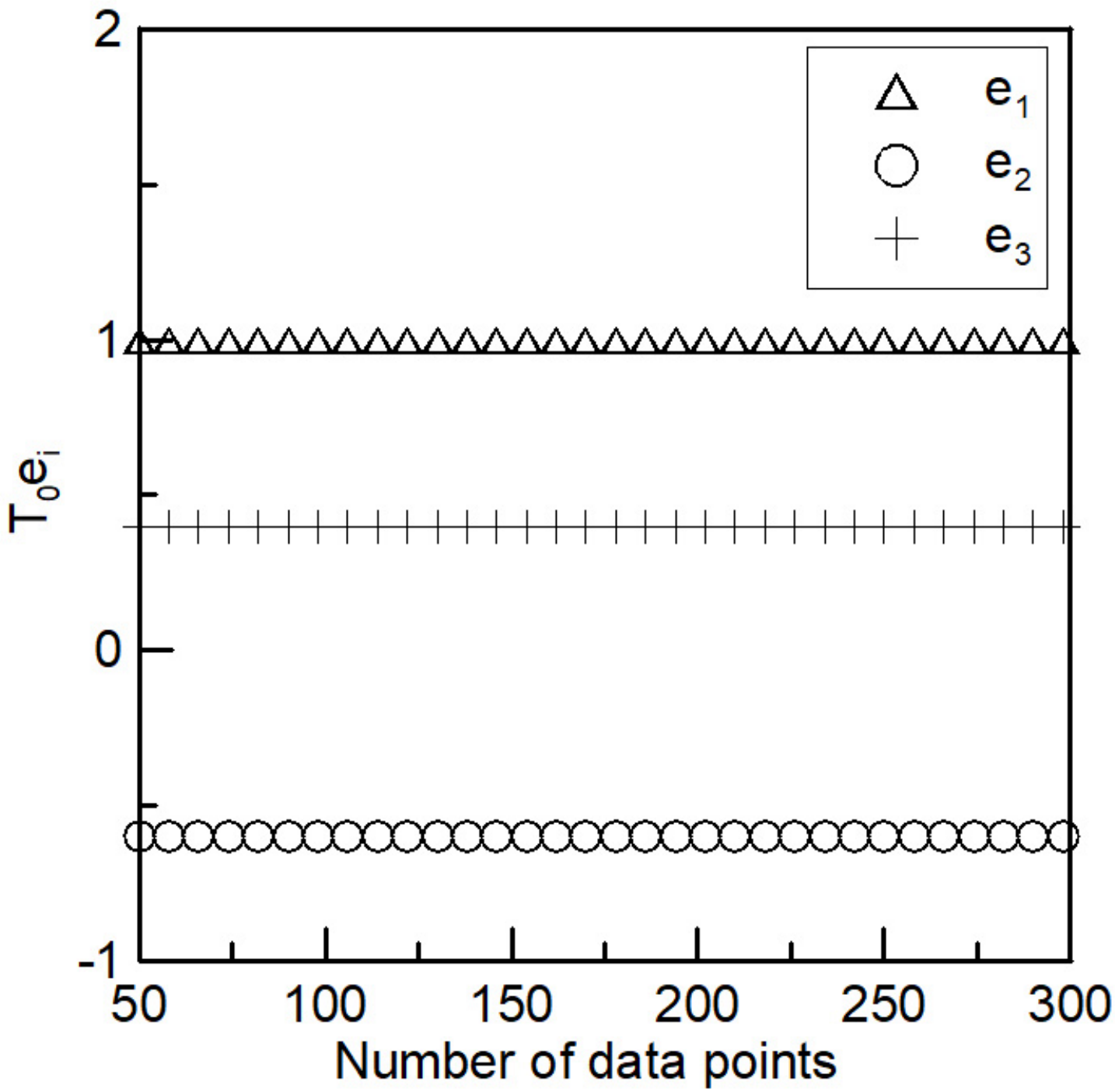}
\includegraphics[scale=0.3]{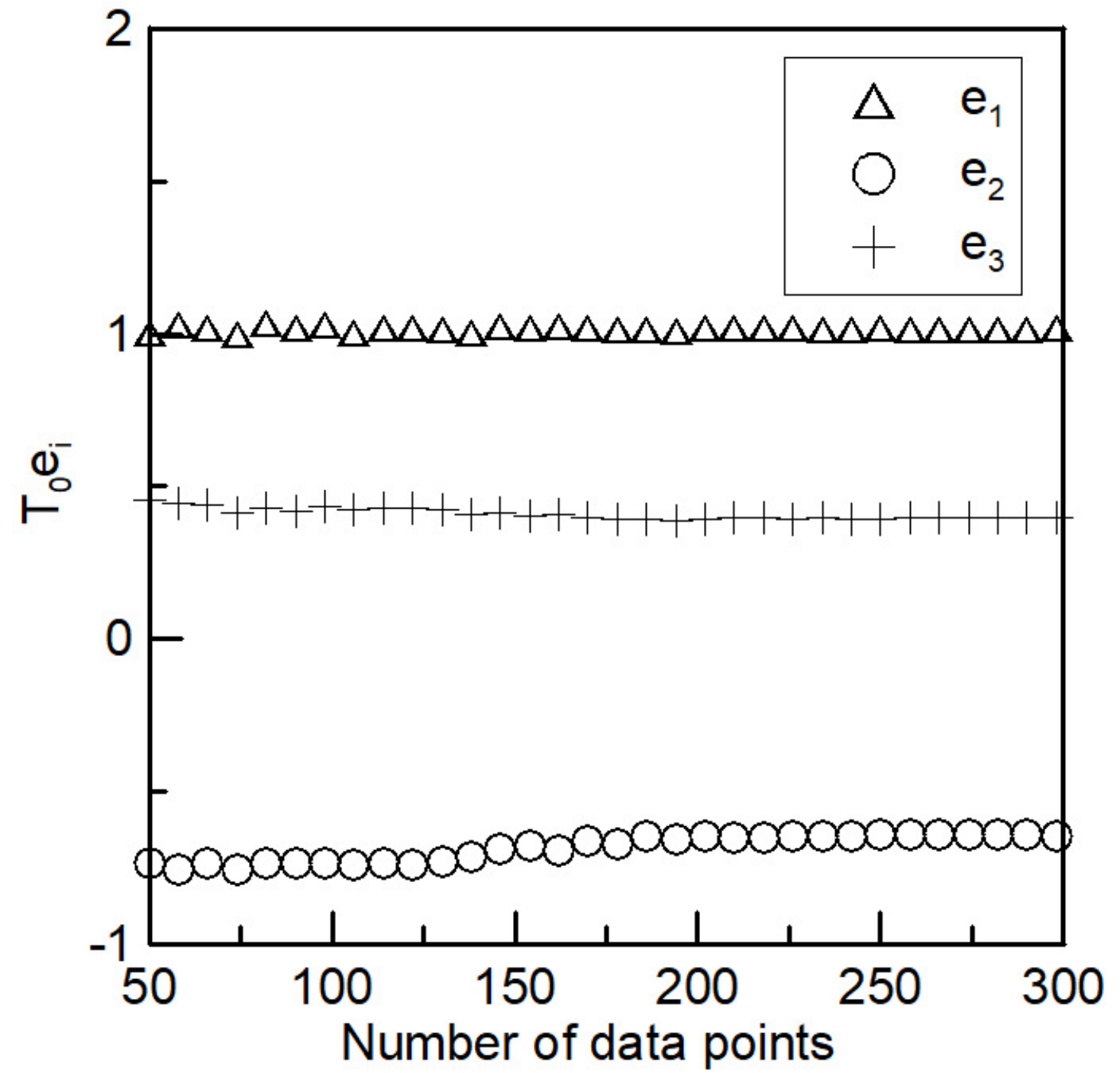} \\
\caption{
Convergence of the Hamiltonian parameter estimation.
The estimated Hamiltonian parameters $e_i$ are plotted against the number of accumulated data points. 
The left and right panels correspond to the estimation results using the clean and noisy Bloch vector trajectories shown in Fig.~\ref{fig:bloch_vec}, respectively. 
The symbols denote the estimated parameters as follows: triangles for $T_0 e_1$, circles for $T_0 e_2$, and crosses for $T_0 e_3$. 
}
\label{fig:parameters_fit1}
\end{center}
\end{figure}

\begin{figure}[h]
\begin{center}
\includegraphics[scale=0.3]{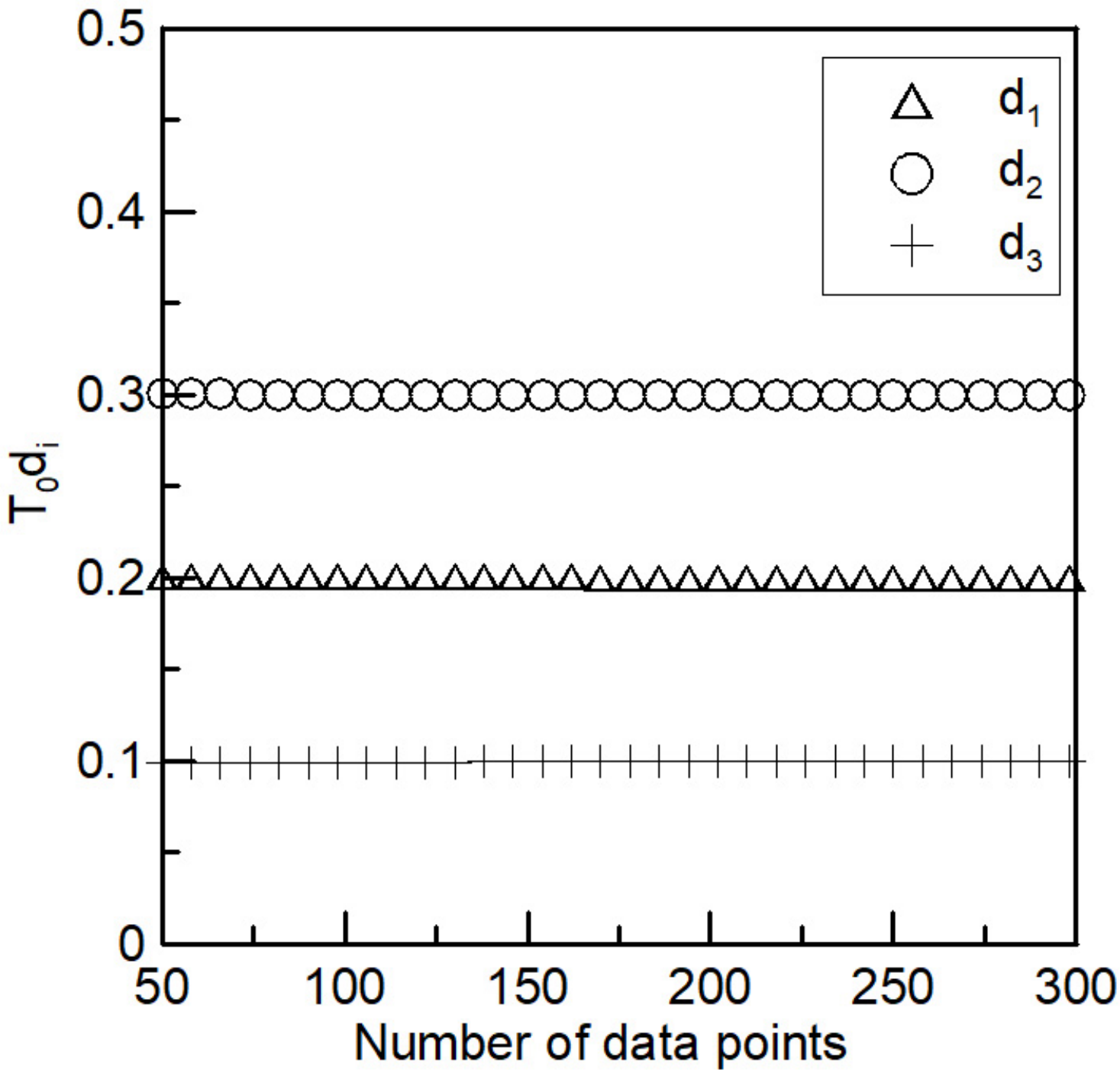}
\includegraphics[scale=0.3]{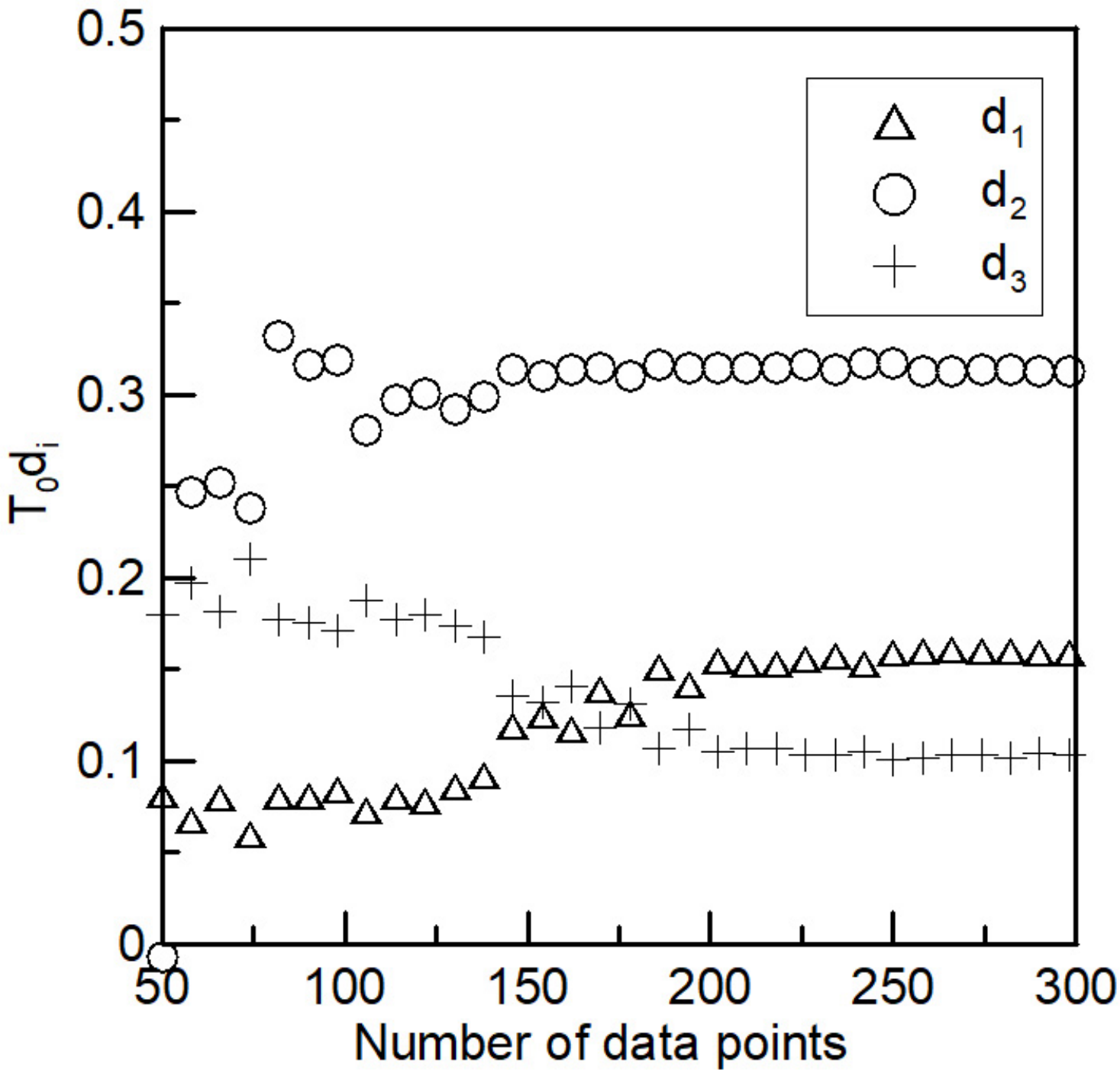}
\caption{
Convergence of the dissipation parameter estimation.
The estimated dissipation parameters $d_i$, corresponding to the relaxation rates, are plotted against the number of accumulated data points. 
Similar to Fig.~\ref{fig:parameters_fit1}, the left and right panels show the estimation using the pure GKSL dynamics and the perturbed data, respectively. 
The symbols denote the estimated parameters as follows: triangles for $T_0 d_1$, circles for $T_0 d_2$, and crosses for $T_0 d_3$. 
}\label{fig:parameters_fit2}
\end{center}
\end{figure}

To validate the proposed estimation method, we performed numerical simulations of a single qubit governed by the unital GKSL equation. 
To prepare a set of input data, we solved the simplified Bloch equation (\ref{eq:gksl_bloch}) with the parameters $\bm{e} = (1.0, -0.6, 0.4)/T_0$ and $\bm{d} = (0.2, 0.3, 0.1)/T_0$, where $T_0$ is a constant with the dimension of time. 
The initial state was set to $\bm{a}(0) = (0.815, -0.007, 0.466)$.
We computed the time evolution for a dimensionless time range from $t/T_0 = 0$ to $10$, and divided this interval into 1000 equally spaced time steps. 
We sampled the data points sequentially from the initial time $t=0$. 
For example, when we state that 100 data points are used, it means that the first 100 consecutive samples from the dataset were utilized. 
These calculations yielded the ideal trajectories shown in the left panel of Fig.~\ref{fig:bloch_vec}. 
To mimic actual experimental conditions, we added small Gaussian white noise to the time evolution. This addition produced the perturbed trajectories shown in the right panel of Fig.~\ref{fig:bloch_vec}.

We then applied our linear regression method to estimate the parameters $\bm{e}$ and $\bm{d}$ from both the clean and noisy datasets.
The performance of our estimation method is demonstrated in Figs.~\ref{fig:parameters_fit1} and \ref{fig:parameters_fit2}. 
Figure \ref{fig:parameters_fit1} shows the estimated Hamiltonian parameters $e_i$ plotted against the number of accumulated data points. 
The left and right panels correspond to the estimation results using the clean and noisy Bloch vector trajectories shown in Fig.~\ref{fig:bloch_vec}, respectively. 
The symbols denote the estimated parameters as follows: triangles for $T_0 e_1$, circles for $T_0 e_2$, and crosses for $T_0 e_3$. 
In the clean case shown in the left panel, the estimates converge extremely rapidly to the true values. 
In the noisy case shown in the right panel, the convergence is slower and exhibits fluctuations, as expected. 
Due to the residual errors, the converged values are slightly shifted from the true values.
However, the effect of noise on these parameters is relatively smaller than its effect on the dissipation rates, as discussed below.

Figure \ref{fig:parameters_fit2} displays the estimated dissipation parameters $d_i$, representing the relaxation rates, plotted against the number of accumulated data points. 
Similar to Fig.~\ref{fig:parameters_fit1}, the left and right panels show the estimation using the pure GKSL dynamics and the perturbed data, respectively. 
The symbols denote the estimated parameters as follows: triangles for $T_0 d_1$, circles for $T_0 d_2$, and crosses for $T_0 d_3$. 
The clean case demonstrates immediate and precise convergence. 
In the presence of noise, we observe large fluctuations in the estimations.
This behavior is related to our initial condition, for which $|\bm{a}|$ is very close to $1$ and the inverse of the BKM metric becomes singular. 
Consequently, any small experimental errors near the initial time significantly diminish the convergence speed of this linear regression approach. 
With a sufficient accumulation of data points, these estimates do slowly converge, although the converged values deviate slightly from the true parameters of the GKSL equation.
Therefore, the incorporation of an error-mitigation procedure is essential for estimating the parameters more precisely within our current scheme.

\section{Discussion and Concluding Remarks} \label{sec:summary}

In this work, we have derived a universal information-geometric identity valid for any single-qubit system and applied it to estimate the parameters of the GKSL master equation from observed time-evolution data. 
This formulation can be regarded as an exact information-geometric extension of TSLs, which are generally formulated as inequalities bounding the rate of state evolution in stochastic and quantum thermodynamics. 
Our mathematical analysis demonstrates that for a single qubit, this inequality saturates into a strict equality, 
because the qubit density matrix belongs to the quantum exponential family where the Pauli matrices serve as sufficient statistics. 
Consequently, the derived identity is a geometric kinematic identity that is satisfied by any smooth mixed-state trajectory, while dynamical information enters only upon the substitution of a specific Bloch equation.

From a practical perspective, this identity enables a linear regression approach to quantum parameter estimation. 
To achieve this, we utilized the BKM metric, which corresponds to the Hessian of the exponential family potential. 
While the SLD metric is commonly used in standard quantum information geometry, the choice of the BKM metric offers a decisive advantage here: it perfectly linearizes the inverse problem. 
Because this linear formulation yields a unique global solution once the full-rank condition is satisfied, it avoids the problematic local minima issues associated with the non-linear optimization methods typically required in standard maximum likelihood estimation.
Our numerical results confirm that the proposed method estimates the Hamiltonian and dissipation parameters with remarkable accuracy under ideal conditions. 
Under the influence of experimental noise, however, the convergence naturally slows down, primarily due to the singular behavior of the BKM metric near pure states. In such cases, the converged values exhibit some deviation from the true GKSL parameters. 
This result highlights the practical importance of introducing robust error-mitigation procedures to estimate the parameters more precisely within our current scheme.

Furthermore, the proposed method provides a natural framework for assessing model validity. 
Because the linear regression imposes no artificial boundary constraints on the parameter space and yields a unique optimal solution, the observation of negative dissipation rates serves as direct, reliable evidence of non-Markovian dynamics or other fundamental deviations from the standard GKSL framework. 
When we combine this geometric approach with appropriate error-mitigation techniques, it offers a powerful and computationally efficient tool for identifying the effects of the environment in Noisy Intermediate-Scale Quantum (NISQ) devices. 
We note that any multi-spin system belongs to the quantum exponential family. Therefore, the mathematical procedure of our information-geometric identity remains valid for higher-dimensional systems, but the required complete operator basis becomes much larger and often impractical.
The application of our method to such systems with more general quantum entanglement is an important task for future research. 
In such cases, we need to clarify how quantum entanglement affects the information-geometric structures and the efficiency of the tomographic estimation.

Additionally, our current framework relies entirely on the Riemannian structure (the metric tensor) of the statistical manifold. 
However, a profound and distinct feature of information geometry is its dualistic structure, which arises from the asymmetry of the divergence and is mathematically captured by the cubic tensor (or dual affine connections) \cite{Amari1985,AmariNagaoka2000,Ay2017}. 
Recent studies have highlighted the physical significance of this asymmetry, relating it to the ``gravitational'' aspects of information in stochastic processes \cite{KoideVenn2025,KoideVenn2026}. 
Interestingly, this cubic tensor is also related to the violation of metric compatibility in differential geometry. 
Such non-metricity may play an essential role in extended theories of general relativity for naturally incorporating the cosmological effects of dark matter and dark energy \cite{KoideVenn2026,Iosifidis2023}. 
From this broader perspective, the study of the cubic tensor has become increasingly important in recent years.
How these dualistic geometric effects encoded in the cubic tensor influence parameter estimation, or how they might provide deeper physical insights connecting open quantum dynamics to generalized theories of gravity, remains a highly non-trivial and intriguing question left for future investigation.

\begin{acknowledgments}
T.K. acknowledges the financial support by CNPq (No.\ 304504/2024-6). A.vdV. gratefully acknowledges funding by the Deutsche Forschungsgemeinschaft (DFG, German Research Foundation) -- Project number 570900169.
A part of this work has been done under the project INCT-Nuclear Physics and Applications (No.\ 408419/2024-5.).
\end{acknowledgments}
\vspace{1cm}

{\bf Declaration of Generative AI and AI-assisted technologies in the writing process}

During the preparation of this work the authors used Gemini (Google) in order to improve the English language expression, verify the mathematical consistency of the derivations, and assist in the development of numerical simulation codes. After using this tool, the authors reviewed and edited the content as needed and take full responsibility for the content of the publication.

\appendix

\section{Derivation of the Bloch Vector Equation with an Affine Term}
\label{app:affine_derivation}

In this appendix, we derive the general equation of motion for the Bloch vector.

The time evolution of the density matrix $\hat{\rho}$ is governed by the GKSL master equation ($\hbar=1$):
\begin{equation}
\frac{d\hat{\rho}}{dt} = \mathcal{L}(\hat{\rho}) = -i[\hat{H}, \hat{\rho}] + \sum_i \gamma_i \left( \hat{L}_i \hat{\rho} \hat{L}_i^\dagger - \frac{1}{2} \{ \hat{L}_i^\dagger \hat{L}_i, \hat{\rho} \} \right) \, ,
\end{equation}
where $\hat{H}$ is a Hamiltonian operator and $\hat{L}_i$ is the Lindblad operator.
Using the Bloch representation $\hat{\rho} = \frac{1}{2}(I + \bm{a} \cdot \bm{\sigma})$, the time derivative of the Bloch vector components $a_i = \mathrm{Tr}[\hat{\rho}\sigma_i]$ is given by:
\begin{equation}
\dot{a}_i = \mathrm{Tr}[\mathcal{L}(\hat{\rho})\sigma_i] = \frac{1}{2} \mathrm{Tr}[\mathcal{L}(I)\sigma_i] + \frac{1}{2} \sum_j a_j \mathrm{Tr}[\mathcal{L}(\sigma_j)\sigma_i] \, .
\end{equation}
This equation can be expressed in the following vector form:
\begin{equation}
\dot{\bm{a}} = \Lambda \bm{a} + \bm{c} \, ,
\end{equation}
where 
\begin{align}
\Lambda_{ij} &= \frac{1}{2} \mathrm{Tr}[\mathcal{L}(\sigma_j)\sigma_i] \, , \\
c_i &= \frac{1}{2} \mathrm{Tr}[\mathcal{L}(I)\sigma_i] \, .
\end{align}

We now evaluate the affine term $\bm{c}$. The unitary part $-i[\hat{H}, I]$ vanishes identically. The dissipative part acts on the identity operator as:
\begin{equation}
\mathcal{L}(I) = \sum_i \gamma_i \left( \hat{L}_i I \hat{L}_i^\dagger - \frac{1}{2} \{ \hat{L}_i^\dagger \hat{L}_i, I \} \right) = 
\frac{1}{2}\sum_i \gamma_i [\hat{L}_i, \hat{L}_i^\dagger] \, .
\end{equation}
Thus, the affine vector is determined by the commutators of the Lindblad operators:
\begin{equation}
\bm{c} = \frac{1}{4} \sum_i \gamma_i \mathrm{Tr}\left( [\hat{L}_i, \hat{L}_i^\dagger] \bm{\sigma} \right) \, .
\end{equation}
If all Lindblad operators are Hermitian such that $\hat{L}_i = \hat{L}_i^\dagger$, the commutators vanish and result in $\bm{c} = \bm{0}$. 
However, for non-Hermitian operators such as the lowering operator $\hat{L} = \sigma_-$, 
the commutator is given by $[\sigma_-, \sigma_+] = -\sigma_z$. 
This lead to a non-zero affine term $\bm{c} \propto \bm{e}_z$ that drives the system toward the ground state. 
See also a similar discussion in Ref.\ \cite{Kaplanek}.

\end{document}